\documentclass[aps,prb,preprint,superscriptaddress,amsmath,amssymb,showpacs]{revtex4}
\usepackage{graphicx}

\begin{document}

 \title{Evidence   of  defect-induced
ferromagnetism in ZnFe$_{2}$O$_{4}$ thin films.}

     \author{C. E. Rodr\'{i}guez Torres}
\affiliation{IFLP-CONICET and Departamento de F\'{i}sica, Facultad de Ciencias Exactas, Universidad Nacional de La Plata, C.C. 67 (1900) La Plata, Argentina}
      \author{F.  Golmar}
       \affiliation{CIC nanoGUNE Consolider
Nanodevices GroupTolosa Hiribidea 76, E-20018 Donostia, San Sebastian - España}
  \author{M. Ziese}
   \affiliation{Division   of  Superconductivity  and  Magnetism,
University  of  Leipzig,  D-04103  Leipzig,  Germany}
  \author{P.  Esquinazi}
 \affiliation{Division   of  Superconductivity  and  Magnetism,
University  of  Leipzig,  D-04103  Leipzig,  Germany}
\author{S.   P. Heluani}
\affiliation{Laboratorio de  F\'{i}sica del S\'{o}lido, Dpto.
de F\'{i}sica, FCEyT, Universidad Nacional de Tucum\'{a}n, Tucum\'{a}n,  Argentina.}

\begin{abstract}
 X-ray  absorption  near-edge and  grazing  incidence
X-ray  fluorescence  spectroscopy  are  employed  to  investigate  the
electronic structure  of ZnFe$_{2}$O$_{4}$  thin  films. The spectroscopy
techniques are used  to determine  the  non-equilibrium cation  site  occupancy as  a
function of depth and oxygen pressure during deposition and its effects on the  magnetic properties. It is found that low deposition pressures below 10$^{-3}$ mbar cause iron superoccupation of tetrahedral  sites without Zn$^{2+}$ inversion, resulting in an ordered magnetic phase with high room temperature magnetic moment.
\end{abstract}

\pacs{61.05.cj, 75.25.-j, 75.47.-m, 75.50.Bb}

\maketitle

\bigskip
\section{Introduction}

Among the functional materials that are being intensely studied to fabricate novel spintronics devices,
the spinel oxides appear as an important alternative to perovskites. Particularly, ZnFe$_{2}$O$_{4}$ can be a half-metallic,
transparent conductor or ferrimagnetic insulator \cite{Yanase:JPSJ:84, Luders:AM06}.
At low dimensions, changing the growth conditions the magnetic and electronic properties of the ferrites can be tuned to
enhance the magnetic moment considerably  and to produce conducting or insulating materials \cite{Ayyappan:APL10}. These properties are very important for the design and fabrication of multilayer structures to produce spin-polarized currents.

It is well known that  bulk ZnFe$_{2}$O$_{4}$ crystallizes in the normal spinel lattice  and that it is  antiferromagnetic with a N\'{e}el  temperature of about 10 K.   Zn$^{2+}$ occupies  tetrahedral (A) and Fe$^{3+}$ octahedral (B)  sites. The interaction between magnetic Fe$^{3+}$  moments is  of superexchange type mediated by  O$^{2-}$  ions and results in the
magnetic  moments   being  antiparallel  to  each   other.  Since  the superexchange interaction  is stronger  for the more  closely situated cations and for the bond  angle closer to 180$^{0}$, the superexchange interaction  of A-O-B  (J$_{AB}$)  is  the strongest,  the  B-O-B (J$_{BB}$)  is weaker, and  that of  A-O-A (J$_{AA}$) is  the weakest.   Accordingly, the magnetic  structure  and   properties  of  spinel-type  oxides  depend strongly   on   the  relative   strength   of  the various   superexchange interactions. In a normal spinel, magnetic  Fe$^{3+}$ ions occupy  only B-sites and  the  weak negative superexchange  interaction  among  Fe$^{3+}$  ions  (J$_{BB}$)  dominates  the magnetic properties of ZnFe$_{2}$O$_{4}$ leading to a low N\'{e}el temperature and consequently,  paramagnetic behavior at room temperature.\\
The situation   is  different,   however,  when   normal   spinel
ferrite  becomes nanosized. Net magnetization at room temperature
can be obtained for nanoparticles obtained by different methods
such as, mechanical synthesis
\cite{Changawa:JPC07,Stewart:PRB07,Shim:PRB06},  solvothermal
method \cite{Blanco:JPC11},  solgel\cite{Shim:PRB06},
co-precipitation \cite{Kamiyama:SSC92, Ayyappan:APL10} and thin
films deposited by sputtering \cite{Nakashima:PRB07} and by pulsed
laser deposition \cite{Chen:JPD08}. Several works  have suggested
that, in this  case,  the ferrite  displays  a nonequilibrium
cation distribution among  the tetrahedral and octahedral  sites
altering its long-range magnetic  ordering. As a consequence  the
magnetic response is drastically enhanced.  It was reported that
when the particle size decreases Fe$^{3+}$ occupancy of both A and
B sites in the nanocrystalline state changes and Zn ions are
transferred from their equilibrium position sites A to B sites.
However, there is still some lack  of clarity  concerning  Zn
non-equilibrium  positions  and their magnetic  effects. It  is
thought  that the  occupation of B-sites by Zn$^{2+}$ brings
Fe$^{3+}$ ions  in both A- and B-sites, and the strong
superexchange interaction  between Fe$^{3+}$ ions in A- and
B-sites forces Fe in B sites to align ferromagnetically  causing
high  magnetization at room temperature \cite{Lotgering:JPCS66,
Ligenza:PSS76, Kamiyama:SSC92, Shim:PRB06,Stewart:PRB07}.  In
addition, several authors reported the clear dependence of
magnetic response on the preparation process, apart from the
particle size \cite{Kundu:PLA03, Shenoya:JMMM04}. Furthermore, it
was found that both,  the magnetic properties and  cation
distribution are extremely dependent  on deposition  conditions
\cite{Chen:JPD08, Takaobushi:APL06}.

Despite progress in the characterization, the  spin configuration
of nanosized  ferrites is  still  unclear and under study
\cite{Kamazawa:PRB03}. As the spinel oxides are ferrimagnets
involving two different sublattices, the resulting magnetic moment
is strongly perturbed by defects in the structural lattice. In
addition, there has been increasing evidence that magnetic order
can be triggered by certain defects in materials that are
nominally non magnetic like certain oxides \cite{Fernandes:PRB09,
Khalid:PRB09, Ramos:PRB10} or pure graphite \cite{Ohldag:NJOP10}.
In a thin film deposition process many of these defects can be
generated. Particularly, it was recently suggested that oxygen
vacancies affect the magnetic properties of synthesized Zn-ferrite
films grown at various oxygen pressures \cite{Sultan:JAP09}. Also
the oxygen  vacancies could be responsible for the semiconducting
behavior found in ZnFe$_{2}$O$_{4}$  thin  films with high
saturation  magnetization  at room  temperature and high Curie
temperature \cite{Chen:JPD08}. There is clearly some need of
employing new experimental procedures using different
techniques that can provide complementary information that
may clarify several open issues in the subject \cite{Makovec:JNR10}.\\

X-ray  absorption near-edge structure (XANES)  technique together
with simulations based on  ab-initio XANES  calculations such  as
code  FEFF8.2  are powerful tools to  identify the modification
of Zn$^{2+}$ and  Fe$^{3+}$ distribution from the equilibrium to
the non-equilibrium
state \cite{Stewart:PRB07, Nakashima:PRB07, Figueroa:JSR09, Akhatar:JOPCM09}. Additionally, Grazing  Incidence X-ray  Fluorescence  (GIXRF) yield the composition, thickness and density  of thin films \cite{Nielsen:PR94, Stoev:SAB99}.\\
XANES  combined  with GIXRF results  in   an  interesting  method
to  relate  thickness-dependent electronic structure with magnetic
properties \cite{Souza:JOP:09}. XANES is not a surface technique
by itself, since the  attenuation length of hard X-rays is a few
micrometres  in any  material.  However, in  the grazing-incidence
geometry near the critical angle  for total reflection, the X-ray
beam is confined within a few  nanometres of the surface. For film
studies, this  confinement has  the  considerable advantage  of
minimizing  the substrate contribution.

In this work, the structural and  magnetic properties of
ZnFe$_{2}$O$_{4}$  films fabricated by pulsed laser deposition
using different  oxygen  partial pressures are reported and
discussed. XANES  spectroscopy combined with GIXRF was used to
determine the non-equilibrium cation site  occupancy as a function
of  depth and its effects on the magnetic properties. Our results
show that the presence of defects as oxygen vacancies or iron
situated on normally non-occupied A sites in the spinel structure
could be the cause for the magnetism in the films.

\section{Experimental}  The ZnFe$_{2}$O$_{4}$  films were  fabricated by
pulsed laser deposition (PLD) from a stoichiometric target onto MgO (001)  substrates. Substrate temperature was 500$^{0}$C  and the oxygen partial  pressure  was varied  between  10$^{-5}$ and  10$^{-1}$
mbar. After deposition the samples  were cooled to room temperature in
vacuum at a  pressure of about 10$^{-7}$ mbar.
  \\ X-ray diffraction measurements employing Cu K$_{\alpha}$ radiation in a Philips X-Pert were made for structural characterization
difractometer.  Scanning Electron Microscopy, Supra 55VP, with  Energy Dispersive Spectroscopy (EDS) was employed  to  study  the  surface morphology and spectral analysis. Magnetization measurements were made in a Quantum Design model MPMS-7 SQUID magnetometer. Magnetic
fields were applied parallel to the films.
 \\ Electrical resistance measurements were performed using a Keithley 2182A Nanovoltmeter and a
Keithley 6221 current source. Electrical contacts were
made with gold wires clenched with indium. We have measured the
resistance with a two-probe techniques. \\
GIXRF measurements were  carried  out  at  the  XRF Fluorescence  beamline  of  the  LNLS
(Campinas, Brazil), using a monochromatic  X-ray beam of 9.7 keV.  The
setup includes 150 $\mu$ m-vertical and 4 mm slits  limiting the beam size
on the sample  mounted on a high precision  goniometer.  Angular scans
around  the  critical  angle   of  total  (external)  reflection  were
performed (between  0 and 2  degrees). The fluorescence  emissions were
collected  using a  15 element  Ge detector.  After the  angular scan,
XANES  Fe  K edge  (7112 eV)  and Zn  K  edge  (9659 eV) spectra  in
fluorescence mode  were collected at different grazing  angles using a
Si (111)  channel-cut monochromator.  The incident beam  intensity and
the energy calibration were monitored using an ion-chamber and a Co (or
Zn) metal standard. The reflected beam and the Co (or Zn) fluorescence
emission were collected using a second ion-chamber and a 15 element Ge
detector, respectively.\\
Room temperature EXAFS (Extended X-ray Absorption Fine Structure) spectra at the
Zn K-edge  were recorded in flourescence mode using a Si(111) monochromator at the XAS2 beamline of the
LNLS (Laboratorio Nacional de Luz Sincrotron) in Campinas, Brazil.  In order to estimate the  X-ray penetration depth
for each incidence angle, the attenuation length  as a function of incidence angle was estimated using the X-ray database
of Lawrence Berkeley National Laboratory \cite{Henke:ADT93}.

\section{Results and discussion}

 X-ray diffractometry indicated  epitaxial growth without any traces of secondary phases.  $\varphi $-scans of the ZnFe$_{2}$O$_{4}$ (511)
refection showed a fourfold symmetry of the film.  Figure 1 shows
the magnetic moment of the films as a function of applied field at
room temperature (a) and at 5 K (b). The diamagnetic signal
 from the substrate was subtracted. The S-shaped M-H curves are evidence for ferrimagnetic order. In addition, a linear high-field response
 in room temperature measurements indicated the presence of a paramagnetic component. This was also deduced from $M$ {\em vs.} $T$ measurements
 (not shown here). The magnetic response diminishes by increasing the oxygen partial pressure from 10$^{-5}$  to 10$^{-1}$ mbar as seen by the
 decreasing of the magnetic moment by more than one order of magnitude. While at room temperature the coercivity is small,
 at low temperatures the films are magnetically rather hard with coercive fields between 447 mT and 1000 mT when deposition pressure varied from P = 10$^{-5}$ to 10$^{-1}$ mBar . The drastic reduction of coercive field at room temperature might indicate the existence of magnetic clusters and a blocking mechanism such as has been observed in references [\onlinecite{Shim:PRB06}] and [\onlinecite{Nakashima:PRB07}] for similar systems.\\

All films are insulating with room temperature resistivity larger than $\rho $= 100~$\Omega $m. Sample ZF02 grown under 10$^{-4}$ O$_{2 }$pressure, had
the smallest electrical resistance and we could measure its behavior under applied magnetic fields between 0.2 and 0.7 T and under light irradiation
between 2.8 and 3.5 eV.  The sample did neither present magnetoresistance nor photoconductivity in the range of fields and energies applied.
SEM micrographs show uniform surfaces in all samples. These results suggest that the sample has  a spatial  uniform magnetic phase at room temperature and that the energy gap in this sample is larger than 3.5 eV.\\
In table I we summarized the results on characterization of samples labeled as ZFO1 to ZFO5.
\begin{table}
    \caption{Measurement parameters of ZFO: Oxygen
partial pressure \textit{p}, film thickness \textit{t}, Fe/Zn ratio,
magnetic saturation moment measured at room temperature and 5 K.} \label{T1}
  \vskip 0.5cm
\begin{tabular}{lccccccc}
\hline  \hline \hline
 sample  & \textit{p} & \textit{t} & Fe/Zn & M(RT)10$^{-4}$& M(5 K)10$^{-4}$ \\
    &(mbar)& (nm)&  &(emu) & (emu)\\
ZFO1 & 10$^{-5}$ & 57& 1.8&1.4&5.0 \\
ZFO2&10$^{-4}$&51   &2.3 &0.2&2.5\\
ZFO3&10$^{-3}$ & 43&1.9& 0.1&1.9\\
ZFO4&10$^{-2}$ & 36&--&--&1.5 \\
ZFO5&10$^{-1}$ &17& 1.8&--&0.3&\\
\hline
\end{tabular}
\end{table}

\begin{figure}
\includegraphics{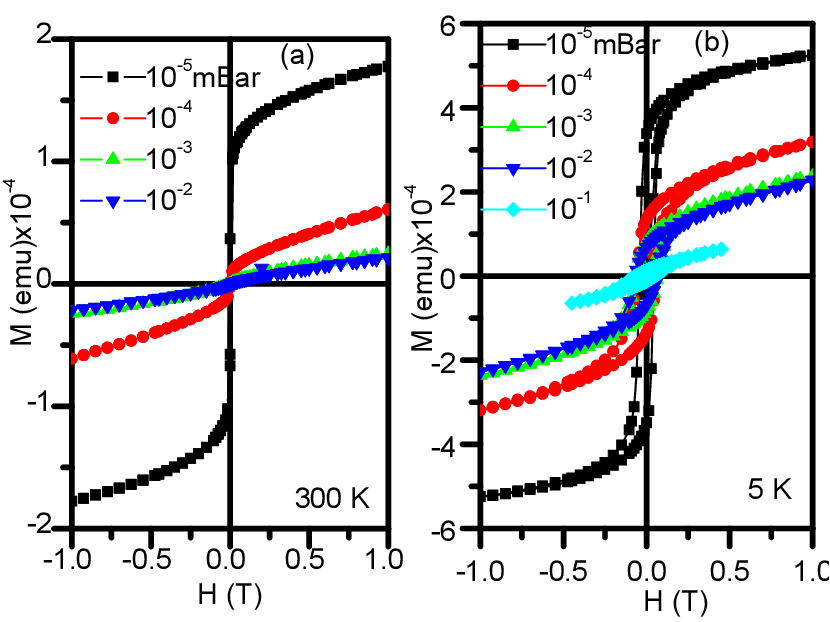}
\caption{Magnetic moment as a function of applied field at room
temperature (a) and at 5 K (b) for the films prepared at different
partial pressures.}\label{mu}
\end{figure}
Using the thicknesses in table I and the magnetic measurements, the estimated magnetization
for sample ZFO1 are 100 emu/cm$^{3}$ and 360~emu/cm$^{3}$ at room temperature and 5~K respectively.  These are large values considering that ZnFe$_{2}$O$_{4}$ with normal spinel  structure is an antiferromagnet.\\
Figure 2  shows the intensity of Fe  and Zn $K_{\alpha}$ lines as a
function  of incident  angle for  all  films. The  intensities are  in
arbitrary  units   and  there  is  not   direct  relationship  between
intensities and concentration. The angle where a jump
in  intensity  was  observed (approximately 0.32 degree) corresponds  to  the  total  reflection
condition.  It  can be  observed  that  Zn/Fe  fraction remains constant  for all  angles indicating  a uniform  distribution  of both
cations  in  depth. Furthermore,  a Zn  enrichment with the increase of  oxygen pressure can be seen:  the curve
corresponding to  Zn almost coincides with  the Fe one  for 10$^{-5}$mbar
and this  is above the  Fe one for  P=10$^{-1}$ mbar. The XRF results were confirmed by EDS (see table I), although it can be seen that the ratio Fe to Zn is close to 2:1 independent of the O$_{2 }$ growth conditions.\\
\begin{figure}
\includegraphics{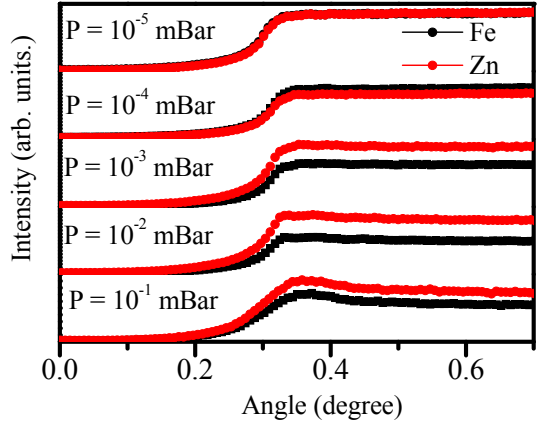}
\caption{Intensity of Fe and Zn $K_{\alpha}$ lines as a function of
incident  angle  for  all  films.  The intensities  are  in  arbitrary
units.}\label{Xanes}
\end{figure}
Figure 3 exhibits Zn (right) and Fe (left) K-edge XANES spectra, corresponding to samples 1, 2 and 5 taken at an incidence angle above total reflection (0.38 degree). The attenuation length for this angle was estimated as 35 nm, and then these spectra contain information from the inner of the film. The spectrum of powder ZnFe$_{2}$O$_{4}$ normal spinel was included for comparison. Zn K-edge spectra of normal spinel have three resolved peaks, A, B and C (indicated in Fig. 1) at around 9664, 9668 and 9672 eV, respectively, a shoulder at around 9677 eV (peak D), plus additional structure at higher energies (peak E). Zn K-edge spectra of samples 1 and 2 have similar characteristics of normal spinel. This indicates that in these films, Zn ions are mostly on A sites. In the case of sample ZFO5 the increase of peak B indicates that there is an important fraction of Zn on B sites \cite{Stewart:PRB07, Takaobushi:APL06, Figueroa:JSR09}.\\
Concerning Fe K-edge, in the case of normal ferrite, the edge position is expected for Fe$^{3+}$ oxidation state and the pre-edge structure is characteristic of Fe in a distorted octahedrically coordinated environment that arises from electronic 1s$\rightarrow$3d quadrupole and 1s$\rightarrow$3d/4p  hybridized orbitals dipole transitions. In XANES spectra of samples 2 and 5, the edge positions are close to ZnFe$_{2}$O$_{4}$ one but the white line amplitude is  smaller which could indicate the presence of Fe$^{3+}$ with a coordination lower than 6 \cite{Stewart:PRB07, Takaobushi:APL06}. Also the pre-peak amplitude are higher than the corresponding to normal ferrite indicating an increase of the degree of orbital p-d mixing that could indicate that the central Fe atoms occupy a more non-centrosymmetric environment.

The decrease of white line and the increase of the pre-peak intensities are larger for samples ZFO1 and ZFO2. Also the shift of the edge to lower energies observed for ZFO1 indicates the presence of Fe$^{2+}$. This sample, which  has the highest magnetic moment, has the lowest white line intensity, the highest pre-peak amplitude and the lowest energy edge.\\
The increase of B peak on Zn K-edge spectrum of ZFO5 could be indicative of inversion in normal spinel, i.e. Zn$^{2+}$ in B sites and Fe$^{3+}$ in A sites.  However, in Fe K-edge spectrum the changes, compared with normal spinel, are minor and this film has the lowest Curie temperature and magnetic moment.
In case of ZFO2 and ZFO5, which are ferromagnetic at room temperature, according
to the Zn K-edge spectrum Zn ions are only in tetragonal A sites then, there is not inversion,
 but the decrease of white line and the increase of pre-peak indicates a decrease of Fe oxygen coordination and probably an increase of Fe in tetrahedral site. \\

In order to study differences between bulk and surface, we present now a comparison between XANES taken with incidence angle below and above  total reflection angle (0.23 degree, attenuation length around 4 nm and 0.38 degree, attenuation length around 35 nm respectively).

\begin{figure}
\includegraphics{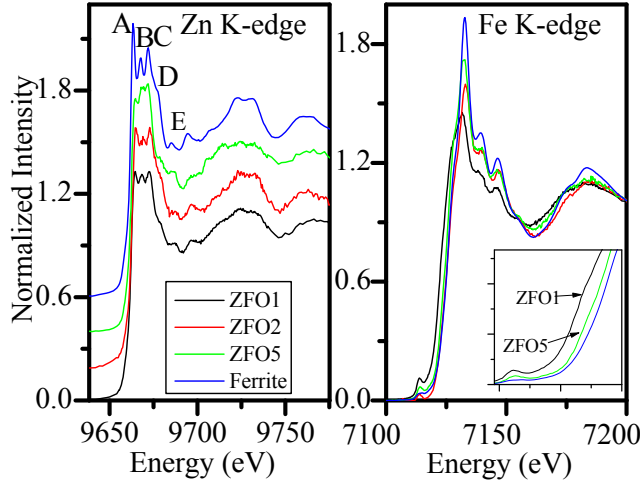}
\caption{(a) Zn and (b) Fe  K-edge XANES using an incident of 0.38 degree, for films deposited at different oxygen  pressures}
\label{spectra1}
\end{figure}

In case of ZFO1 (Figure 3),  no major changes were observed in the Zn K-edge. The
decrease  of peak C could be related to the elimination
of multiple scattering paths due to the high contribution of surface atoms.
The same can be observed for ZFO2 (see Figure 4). In Fe K-edge,
the changes that evidence an increase of Fe in A sites and/or vacancies in
 the B site octahedron are less pronounced at the  surface. In the case of ZFO5 (Figure 4)
  the tendency is similar but the difference with normal spinel is smaller.\\

  \begin{figure}
\includegraphics{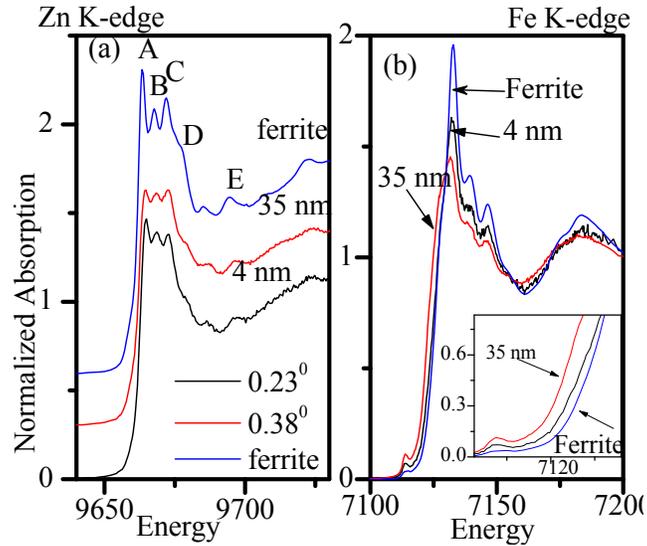}
\caption{(a) Zn and (b) Fe  K-edges XANES, taken for film deposited at 10$^{-4}$ mbar for two different incidence angles.}
\label{spectra2}
\end{figure}
\begin{figure}
\includegraphics{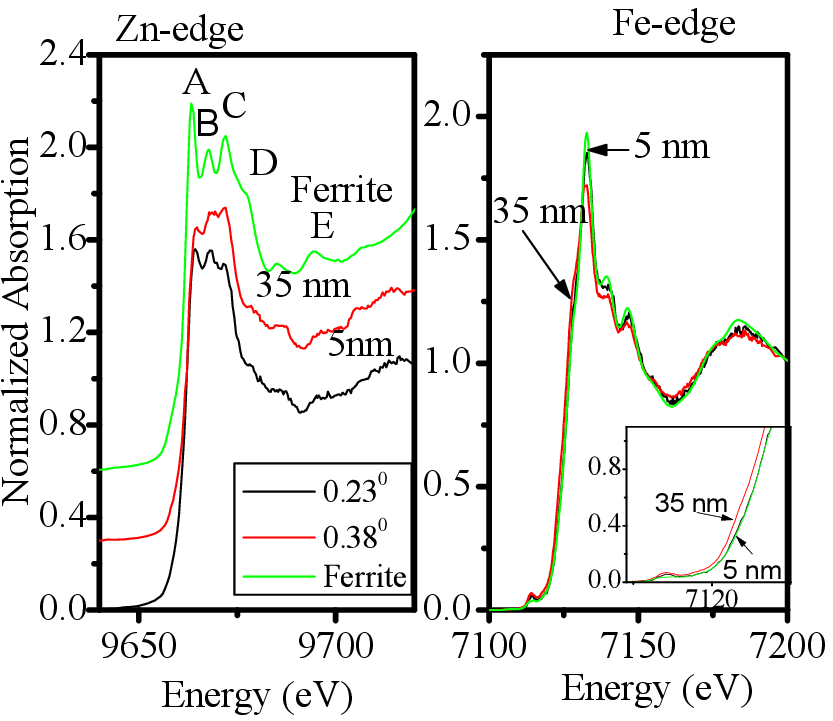}
\caption{(a) Zn and (b) Fe  K-edge XANES, taken for film deposited at 10$^{-1}$ mbar for two different incidence angles.}\label{Spectra3}
\end{figure}
Regarding XANES results, we can conclude that in the films grown at low pressure Zn$^{2+}$ ions are in A sites (coordination four), and   Fe$^{3+}$ are octahedrically (B site) and tetrahedrically coordinated (A site). Also, there is fraction  Fe$^{2+}$ probably in B sites such as in Fe$_{3}$O$_{4 }$ structure. The increase of Fe$^{3+}$ in A sites and Fe$^{2+}$ is favored by the deposition at low oxygen pressure and then due probably to the oxygen vacancy formation. A probe of that is the decrease of features in the  XANES Fe K-edge taken with incidence angles lower than total reflection one, i.e. corresponding to the surface region where available environment oxygen neutralize oxygen vacancies. Then, in the case of low pressure growth films there is an increase of Fe$^{3+}$ tetrahedral sites without inversion.

 This significant result is supported by EXAFS. Figure 6 shows the Fourier transform (FT) of oscillations extracted from Zn k-edge  using Athena program of ZFO1 and ZFO5 films compared with FT of bulk ZnFe$_{2}$O$_{4 }$  one.

FT of sample ZFO1 is qualitatively equal to bulk ZnFe$_{2}$O$_{4 }$, reasserting that Zn atoms are in the tetrahedral sites. In the case of ZFO5, the presence of an additional peak around 2.6 A is an indicative that there are Zn atoms at octahedral sites \cite{Stewart:PRB07}.

   \begin{figure}
\includegraphics{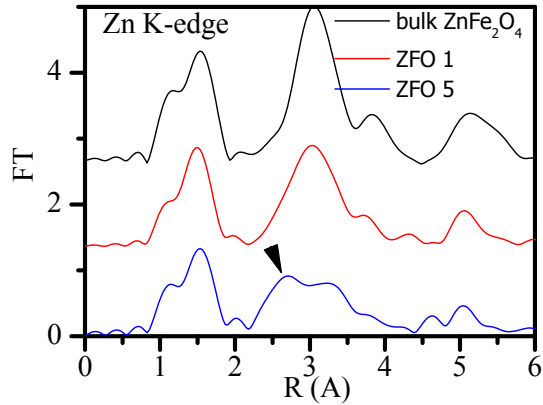}
\caption{ Fourier transforms of k$^{2}\chi$(k) oscillations (Zn K-edge) of bulk ZnFe$_{2}$O$_{4 }$, ZFO1 and ZFO5 film ferrites . Arrow indicates the new coordination appear in FT of sample ZFO5 indicative that there are Zn atoms at B sites.}
\label{FTZn}
\end{figure}

The situation is different to the scenario found in references \cite{Stewart:PRB07} and \cite{Nakashima:PRB07}: when Zn occupancy of 8-sites increases, Fe  decreases, indicating some degree of inversion and magnetic response enhanced with such inversion. In our case the increase of Fe$^{3+}$  in A sites is not due to Zn and Fe ions inversion, but the mechanism that explains the increase in the magnetization is probably the same as the one reported for inverse spinel. The increase of Fe$^{3+}$  tetrahedrically coordinated (A site) increases Fe$^{A}$-Fe$^{B}$ pairs antiferromagnetically coupled by superexchange, forcing Fe$^{B}$-Fe$^{B}$ pairs to align ferromagnetically and  also because Fe individual magnetic moments probably no longer cancel. The experimental results report in this work suggest that the increase of Fe$^{3+}$ in A sites is not due to Zn and Fe ions inversion but of iron situated on normally non-occupied A sites in the spinel structure. In this structure only one eighth of the A sites in a unit cell are actually occupied and so some iron ions at these normally unoccupied sites will act as impurities and will not influence the crystallographic properties.\\
 One more aspect in which oxygen pressure could influence is that as the magnetic interaction in the spinel is an indirect interaction, missing oxygen gives rise to a variation of exchange fields for the ions and a spin/glass like state is formed. Also, the reduction  of Fe$^{3+}$ ions into Fe$^{2+}$  ions located in octahedral sites would strength ferromagnetic coupling between B-B sites such is the case in magnetite. In the case of Fe$^{2+}$  located on the A sites there is an increase of magnetic moment due to the antiferromagnetic coupling with Fe$^{3+}$ but the change will not be as big as for anti-sites.

In summary, we have shown that ferrites grown under low O$_{2}$
pressure conditions  have a large magnetic moment. We found that
the inversion mechanism is not responsible for the enhancements of
the magnetic interaction -- as was found in similar systems -- but
the presence of defects as oxygen vacancies or iron situated on
normally non-occupied A sites in the spinel structure. We have
also shown that controlling the oxygen pressure during the
deposition, it is possible to obtain conductive or insulating Zn
ferrites. These findings allow to control the magnetic and
electric transport  properties of spinel ferrites controlling the
oxygen concentration.

\begin{acknowledgments}
We thank Dr. Silvana Stewart for fruitful discussions and Dr. Azucena Mudarra Navarro for collaboration in XAFS experiments.\\
 This work  was partially  supported by Laboratorio Nacional de Luz Sincrotron, Campinas-Brasil; by CIUNT under Grants 26/E439 by ANPCyT-PICTR 20770 and 35682, by the German-Argentine PROALAR Grant Nr. D/08/11707 and the Collaborative Research Center SFB 762
"Functionality of Oxide Interfaces"
\end{acknowledgments}


\end{document}